# Ring Pearcey vortex beam dynamics through atmospheric turbulence

SHAKTI SINGH,[1] SANJAY KUMAR MISHRA,[2] AKHILESH KUMAR MISHRA[1,3],*

[1]Department of Physics, Indian Institute of Technology Roorkee, Roorkee-247667, India
[2]Photonics Research Lab, Instruments Research and Development Establishment, Dehradun-248008, India
[3]Centre for Photonics and Quantum Communication Technology, Indian Institute of Technology Roorkee, Roorkee- 247667, Uttarakhand, India
*Corresponding author- akhilesh.mishra@ph.iitr.ac.in

**Abstract:** The subject area of free space optical communication (FSO) with optical beam carrying orbital angular momentum (OAM) has attracted a great deal of research attention since last two decades. Efforts to understand, model and execute communication links through turbulent atmosphere with OAM beams have gained particular importance. In this regard, different types of shape preserving beams, which can withstand turbulences of varying strengths, have been proposed and studied. In this paper, we present a numerical investigation of the propagation characteristics of ring Pearcey vortex beam (PVB) through turbulent atmosphere. The study details on both moderate as well as strong atmospheric turbulences. Modified von Karman model has been relied on to model random phase screen. In moderate turbulence, the ring PVB preserved its singularity. In strong turbulence, the ring PVB preserved its singularity for short propagation distances but lost its singularity at longer propagation distances. We found that upon increasing the value of topological charge ($l$), aperture averaged scintillation index (SI) increases. We calculated the aperture averaged SI for different truncation factors and noticed that the ring PVB with a truncation factor $b = 0.1$ performed better in stronger turbulence. In moderate turbulence, the aperture averaged SI performed better for shorter propagation distances and relatively larger truncation factors. Further, we calculated the aperture averaged SI for spatially chirped ring PVB, and it has been found that aperture averaged SI improved largely for negatively chirped ring PVB. Furthermore, on comparing the aperture averaged SI of ring PVB and ring Airy vortex beam (AVB), it has been noticed that in strong turbulence ring PVB exhibited better aperture averaged SI. Additionally, we have calculated the beam wander for ring PVB and ring AVB and found that ring PVB demonstrates better beam wander.

## 1. Introduction

Optical beam propagation through atmosphere has drawn considerable attention due to its applications in many areas such as free space optical communication (FSO) [1], remote sensing [2], laser guided defence system and active imaging [3]. Turbulence strongly affects the refractive index of the atmosphere and therefore it gives unexpected consequences in the optical beams such as intensity fluctuation, beam wandering, decrease in spatial coherence and loss of optical power [4-6]. Intensity fluctuation in optical beams is characterised by a physical quantity called scintillation index (SI) [7]. Scintillation results due to the distortion in the wavefront of optical beam by turbulence. In FSO, the SI of an optical beam at the detector end should be minimum. Beam wander is another important parameter in FSO. It is described by variance of displacements of beam centroid [7]. An optical beam that has ability to resist the turbulence strongly is preferred for FSO. Nondiffracting and self-healing beams such as Bessel [8] and Airy beams [9-11] have the potential to overcome turbulence induced deformations.

Due to rigorous stochastic nature of atmospheric turbulence, analytical solution of Helmholtz equation exists only for limited number of optical beams. These solutions are restricted to the lower order statistics. Therefore, numerical modelling becomes essential tool for understanding the nature of the complex optical beam through the atmospheric turbulence.

Scintillation of the optical beams propagating through the turbulence have extensively been studied for scalar beams such as Laguerre-Gauss beam [12], Bessel beam [13], Airy vortex beam [14], cos and cosh-Gaussian beam [15], elliptical vortex beam [16] and flat top beam [17]. We would like to note here that vortex beam is a special kind of beam, which possesses undefined phase at the centre of the beam [18-20]. To overcome the turbulence induced intensity fluctuations many techniques have been proposed. Coherence of optical beams also affect the SI and it has already been reported that partially coherent beams exhibit lower SI compared to the fully coherent beams [21-23]. A coherent beam with non-uniform polarization also shows lower SI compared to the coherent beam with uniform polarization [24-27]. Beam arrays, which are constituents of spatially separated beamlets, have been used for reducing the scintillation by adjusting the spatial separation among the beamlets [28-30]. Any deviation in the spatial separation among beamlets causes significant increase in the SI. Dynamics of optical beam is strongly affected by initial defining parameters of the optical beam. Spatial chirp is one such parameter that strongly affect the dynamics of optical beam. The effects of spatial chirp on dynamics of ring Pearcey Gaussian vortex beam and Airy Gaussian vortex beam in non-turbulent media have also been investigated [31-33].

Current times are witnessing growing interest in FSO and remote sensing using optical vortex beams. Vortex beams carry orbital angular momentum (OAM), which can be multiplexed to encode information data for transmission in FSO [34]. Vortex beam propagation through atmospheric turbulence has also been explored from the perspective of quantum communication and quantum entanglement [35-36].

During last decade, abruptly autofocusing (AAF) beams have emerged as another interesting class of beams that exhibit unprecedented properties like nondiffracting and self-healing [10]. AAF beam have important applications in micromanipulation and biomedical treatment. Circular airy beam is the first AAF beam which was introduced in 2010. The auto focussing behaviour of such beams owes its origin in the optical field structure [9-10]. Circular Pearcey beam is another example of AAF beams, whose analytical expression comes from transforming the Pearcey function to the cylindrical coordinate [37-39]. Optical vortex when assigned to any optical beam modifies its characteristics extensively. AAF beams superimposed with optical vortices have also been explored extensively in the recent literature. In particular, propagation characteristics of AAF airy beam with optical vortices has lately been investigated in [14]. AAF Pearcey beams have also been studied experimentally long back [39]. During propagation in atmospheric turbulence AAF beams balance the beam wander and spreading of beams, which in turn helps in reducing the crosstalk talk, SI and bit error rates, which ultimately improves the FSO link.

In the present work, we numerically report the propagation characteristics of ring Pearcey vortex beam (PVB) through atmospheric turbulence. The propagation of ring PVB through atmospheric turbulence is modelled using modified Von Karman spectrum with FFT method. The paper is arranged as follows- in section (2) we have elaborated the numerical model. In section (3), we have discussed the profile of ring PVB and the associated numerical simulation parameters. Section (4) discusses the intensity variation of ring PVB in free space and turbulent atmosphere with varying strength of turbulence. The variation of aperture averaged SI with propagation distance for different topological charge and truncation factor are also elaborated in this section. To the best of our knowledge, no work on the spatially chirped optical beam through atmospheric turbulence is reported earlier. This section fills this gap and entails investigation on propagation dynamics of the aperture averaged SI of differently chirped ring PVB in turbulent atmosphere. Further, the aperture averaged SI of ring PVB is compared with that of ring AVB. Section (5) discusses the beam wander of ring PVB and ring AVB. Finally, section (6) concludes the work.

## 2. Numerical model

Optical beam propagation through any medium is modelled using Helmholtz equation. In turbulent atmosphere due to the random nature of refractive index, propagation is modelled instead using stochastic Helmholtz equation. Randomly varying refractive index is characterized by power spectral density (PSD). PSD describes the statistical distribution of the number and size of turbulent eddies [4]. Optical beam propagation through atmospheric turbulence is modelled using multiple phase screen method [40]. In this method, we treat the turbulent atmosphere as a collection of thin random phase screens placed along the propagation direction with equal interval of free space. Randomly varying phase imposed on these phase screens enables them to mimic atmospheric turbulence. Different types of PSDs have been proposed for modelling the propagation of optical beam through atmospheric turbulence. The widely employed one is Kolmogorov PSD, whose mathematical expression is given as [7]

$$\phi_n(\kappa) = 0.033 C_n^2 \kappa^{-11/3} \quad \text{for } \frac{2\pi}{L_0} < \kappa < \frac{2\pi}{l_0}. \tag{1}$$

In above PSD in the limit $\kappa \to 0$, it contains a singularity in the form of non-integrable pole. Another PSD model called Tatarskii spectrum was proposed, which is given by [7]

$$\phi_n(\kappa) = 0.033 C_n^2 \kappa^{-11/3} exp\left(\frac{-\kappa^2}{\kappa_m^2}\right) \text{ for } \kappa > \frac{2\pi}{L_0}, \tag{2}$$

where $\kappa_m = \frac{5.92}{L_0}$. This spectrum also contains a singularity in the limit $\kappa \to 0$. Another important PSD called von Kármán spectrum has also been put forth, which is expressed as

$$\phi_n(\kappa) = 0.033 C_n^2 (\kappa^2 + \kappa_0^2)^{-11/6} \qquad \text{for } 0 \le \kappa \ll \frac{2\pi}{l_0}. \tag{3}$$

Generally, the refractive index spectrum follows the same spectral law as that of temperature. However, none of the above spectrum show this property. Hill developed a theoretical spectrum for temperature fluctuations in the atmosphere whose approximate analytical expression is given by [7],

$$\phi_n^M(\kappa) = 0.033 C_n^2 \left[1 + 1.802\left(\frac{\kappa}{\kappa_l}\right) - 0.245\left(\frac{\kappa}{\kappa_l}\right)^{7/6}\right] \frac{exp(-\kappa^2/\kappa_l^2)}{(\kappa^2 + \kappa_0^2)^{11/6}}, \qquad \text{for } 0 < \kappa < \infty, \kappa_l = \frac{3.3}{l_0} \tag{4}$$

this spectrum is also known as modified atmospheric spectrum.

For taking the entire range of $\kappa$ a modified Von Karman type PSD was proposed. The expression for Von Karman type PSD is given by [12]

$$\phi_n(\kappa) = 0.033 C_n^2 (\kappa^2 + \kappa_0^2)^{-11/6} \exp\left(\frac{-\kappa^2}{\kappa_m^2}\right), \tag{5}$$

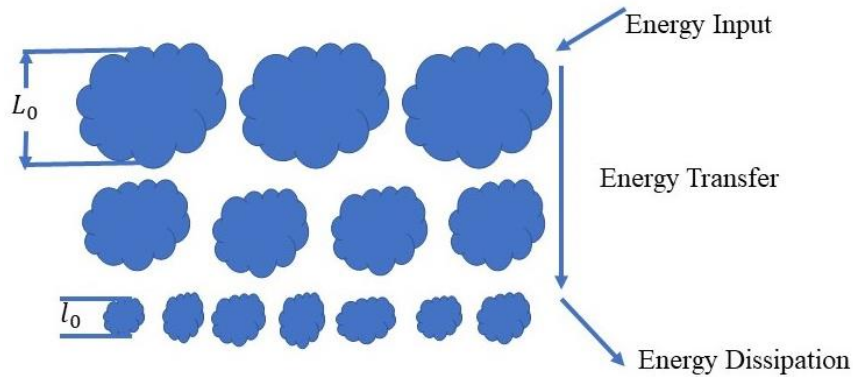

Fig. 1. Energy cascade theory of turbulence [7].

where $C_n^2$ represents the refractive index structure constant that represents the strength of atmospheric turbulence. The typical value of $C_n^2$ varies from $10^{-17} m^{-2/3}$ to $10^{-12} m^{-2/3}$ [12]. $\kappa$ represents the magnitude of three-dimensional spatial frequency vector in cartesian coordinate. In the above equation, $\kappa_0 = \frac{2\pi}{L_0}$ and $\kappa_m = \frac{5.92}{l_0}$, where $L_0$ and $l_0$ represent the outer and the inner scales of the atmospheric turbulence respectively. In turbulent atmosphere, unstable air masses due to influence of inner forces break up into smaller turbulent eddies to form a continuum of eddies for transfer of energy from outer scale $L_0$ to inner scale $l_0$. Atmosphere behaves like isotropic and homogeneous only when the size of turbulent eddies lies between inner scale $l_0$ and outer scale $L_0$. The typical magnitude of inner scale is of the order of milli-meter while that of outer scale is of the order of meter near the ground and their values increase as we move upward. The number of random phase screens required to model the turbulence depends upon the propagation distance and strength of the turbulence. In the present numerical investigation, we have employed the widely used modified Von Karman type PSD as given in equation (5). Inner and outer scale values are considered to be $l_0 = 1 cm$ and $L_0 = 3\ m$ respectively [12]. We have used 20 random phase screens at equal interval of the total propagation distance, which is $2\ km$ in our numerical experiment.

The random phase screen is generated by considering an $N \times N$ array of complex Gaussian random number $a + ib$ and then this complex number is multiplied with the square root of phase spectrum. The phase and power spectrums are related and are expressed as [12]

$$\phi_\theta(\kappa) = 2\pi k^2 \delta z \phi_n(\kappa), \tag{6}$$

where $k$ is the wavevector, $\delta z$ is the distance between the phase screens. Now we multiply the complex Gaussian random number by $\Delta_\kappa^{-1}\sqrt{\phi_\theta(\kappa)}$, where $\Delta_\kappa^{-1} = 2\pi/N\Delta$, $N$ is the number of sampling point and $\Delta$ is the spatial sampling interval. The result is then inverse Fourier transformed to get the random phase $\theta_1 + i\theta_2$. The generation of phase screen by above method for modelling the propagation through the atmospheric turbulence is called FFT method. Throughout our numerical investigation we have chosen $\theta_1$ as random phase screen. To validate our numerical model, we have first generated the results of published in reference [12]. Propagation chart (algo) of optical beam through atmospheric turbulence is illustrated in figure (2). First, we propagate the optical beam up to a distance $\delta z$ under diffraction by employing angular spectrum method, where 2D Fourier transform of the optical field is multiplied with free space transfer function, whose expression is given by the equation (7). The resultant optical field illuminates the phase screen, and this process continues until the final distance is reached.

$$A(f_x, f_y) = \exp\left(i\delta z \sqrt{k^2 - 4\pi^2 (f_x^2 + f_y^2)}\right). \tag{7}$$

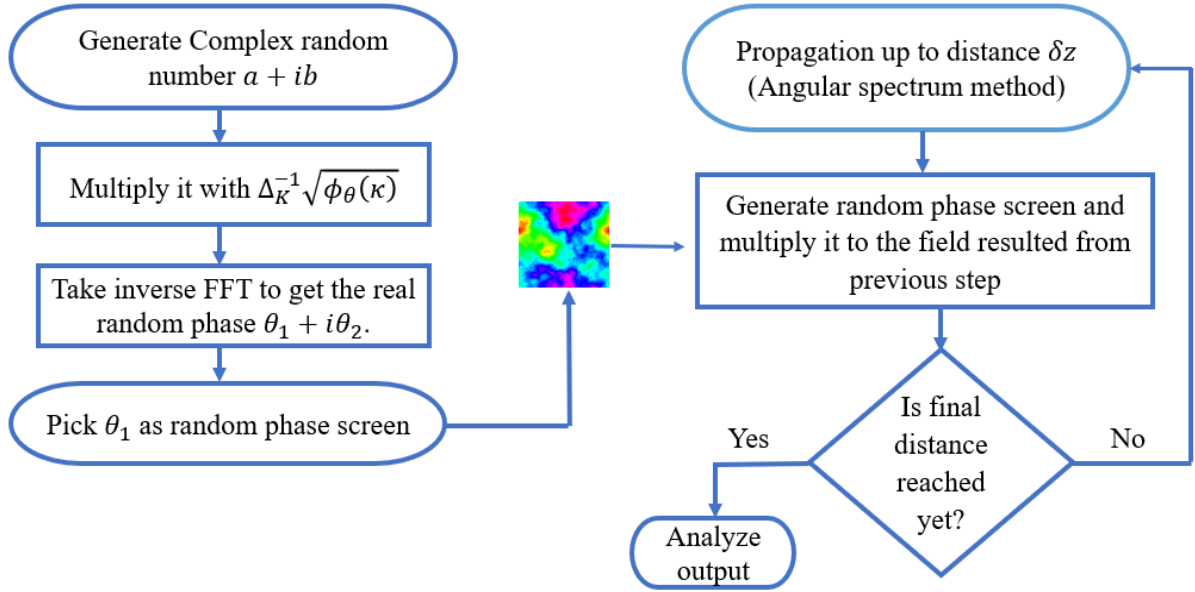

Fig. 2. Propagation flow chart used in our numerical model.

## 3. Simulation parameters

In our numerical investigation, we have studied the propagation of ring PVB through atmospheric turbulence whose complex field distribution at the input is given by

$$E(r, \phi, 0) = Pe\left(\zeta_0, \frac{-r}{w_0}\right) exp\left[b\left(\frac{-r}{w_0}\right)\right] exp(il\phi), \qquad (8)$$

where $Pe$ is the Pearcey function, which is defined by an integral representation as given by

$$Pe(x, y) = \int_{-\infty}^{\infty} \exp\left(is^4 + is^2y + isx\right) ds, \qquad (9)$$

here $x$ and $y$ are dimensionless transverse variables. In equation (8), $w_0$ is the width of the primary ring of the beam, $l$ is the topological charge of ring PVB, $b$ is truncation factor and $\phi$ is the azimuth angle $\zeta_0$ denotes a constant. In our simulation, we have taken $w_0 = 3\ cm$, wavelength $\lambda = 2\ \mu m$, $l_0 = 1\ cm$, $L_0 = 3\ m$, $N = 500$ $\zeta_0 = 0$ and rest of the parameters are specified in respective figure captions. We have propagated the optical beam up to $2\ km$ and used $20$ random phase screens at an interval of $100\ m$. The simulation has considered 500 independent realizations to provide sufficient statistics for the calculation of averaged irradiance and SI.

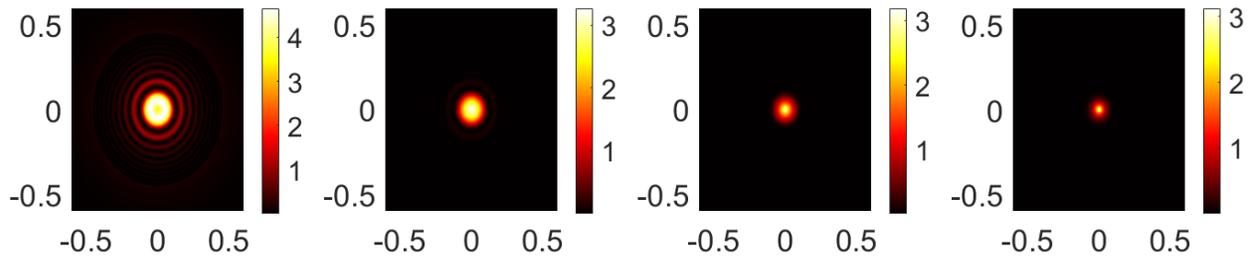

Fig. 3. Input beam intensity profile of ring PVB with $l$=1 for different values of truncation factor, (a) $b$=0.1, (b) $b$=0.3, (c) $b$=0.5 and $b$=0.7 in (d). (Horizontal and vertical axes are in meter ($m$))

## 4. Results and discussion

### 4.1 Propagation effects in atmospheric turbulence

To understand the dynamics of ring PVB in the turbulent atmosphere, we, in figure 4, have plotted the evolution of intensity profiles of ring PVB in both moderate and strong turbulences. The first row, in figure (4), shows the evolution of ring PVB in free space without turbulence, wherein the optical beam demonstrates self-focusing. Second row of the figure shows the beam evolution in moderate turbulence ($c_n^2 = 10^{-14}$), wherein we observe that although the ring PVB maintains its singular characteristic, its self-focussing strength has got weakened (observe the difference in colour bars). The singular characteristic as preserved by ring PVB is due to its self-healing nature. For strong turbulence ($c_n^2 = 10^{-12}$), as depicted in the third row of figure (4), we observe that despite the strong turbulence ring PVB tries to preserve its singularity up to $500m$ but on furthering the propagation the ring PVB loses its typical singular behaviour. We also observe that in strong turbulence ring PVB could not preserve its self-focussing property. For smaller distances, the effect of atmospheric turbulence is very moderate but on increasing the distance the beam gets distorted heavily and irreparably.

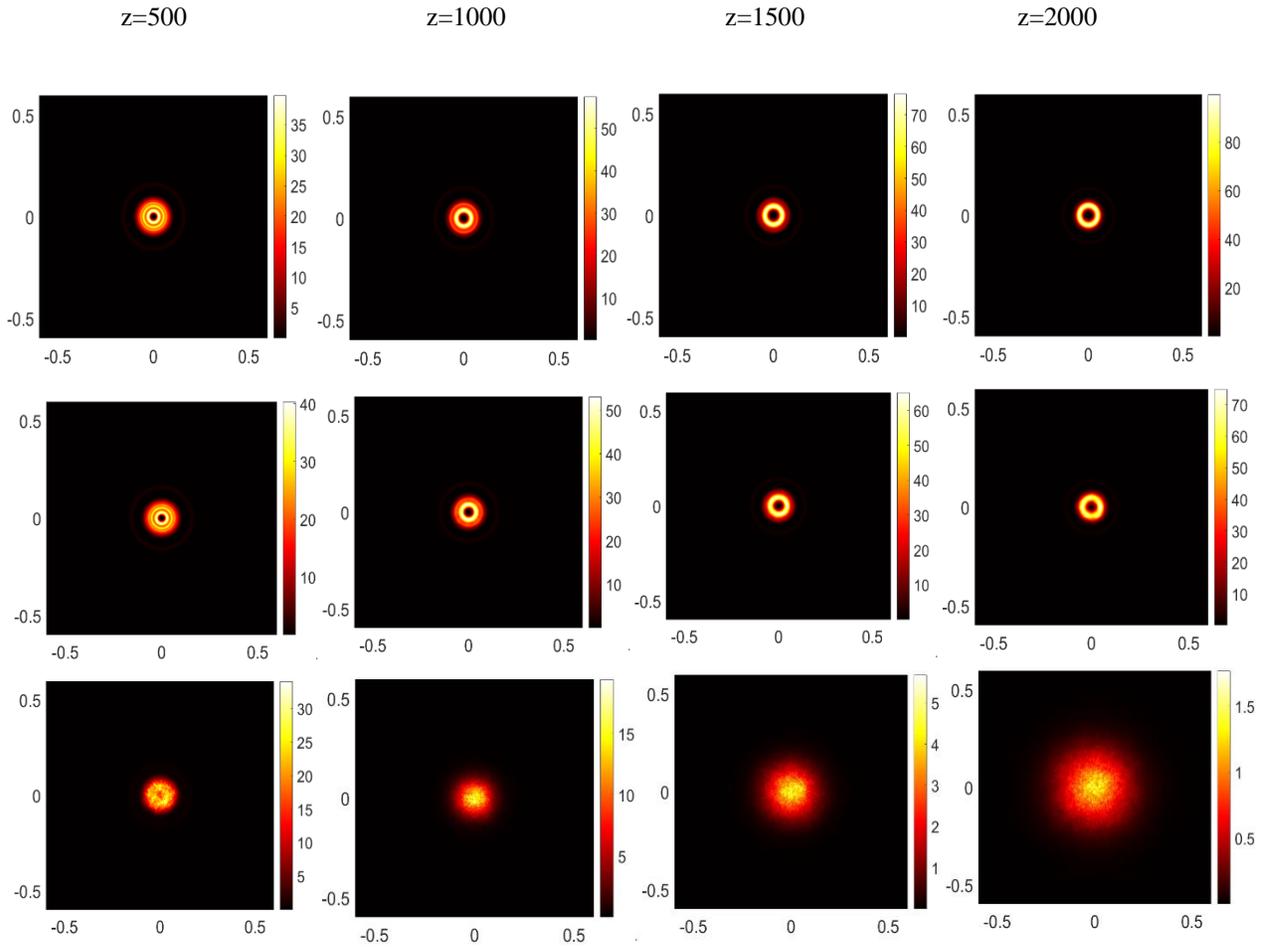

Fig. 4. Transverse intensity profile evolution of ring PVB for $l = 1$ and $b = 0.1$ in free space without turbulence (1st row), with moderate turbulence ($C_n^2 = 10^{-14}$) (2nd row) and with strong turbulence ($C_n^2 = 10^{-12}$) (3rd row). The four columns represent the transverse intensities at z=500 $m$, z=1000 $m$, z= 1500 $m$ and at z=2000 $m$ respectively. (Horizontal and vertical axes are in meter ($m$))

## 4.2 Aperture averaged scintillation in atmospheric turbulence

The optical beam quality through atmospheric turbulence is characterized by on-axis SI. The SI at a transverse position $(x, y)$ in detector plane is defined by [7]

$$\sigma_I^2(x, y, z) = \frac{\langle\langle I(x,y,z)^2\rangle\rangle}{\langle I(x,y,z)\rangle^2} - 1, \qquad (10)$$

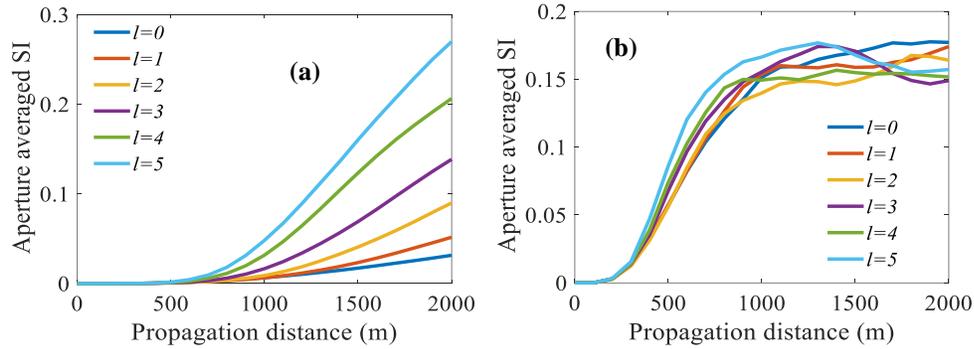

Fig. 5. Aperture averaged SI for ring PVB with propagation for moderate (a) $C_n^2 = 10^{-14} m^{-2/3}$ and strong (b) $C_n^2 = 10^{-12} m^{-2/3}$ turbulence parameters.

where $I$ is the irradiance of the optical beam and $<>$ represents the ensemble averaging. $\langle I(x,y,z) \rangle$ is obtained by summing the independent realizations and then dividing the outcome by total number of realizations. Since the detector, used for measuring the SI in experiments, have finite aperture size, it gives us aperture averaged effects. To align our numerical results with the experimental measurements, we calculate the aperture averaged SI, which is given by [14]

$$S(z) = \frac{\left\langle \left( \int_{-R}^{R} \int_{-R}^{R} I(x,y,z) dx dy \right)^2 \right\rangle}{\left( \int_{-R}^{R} \int_{-R}^{R} I(x,y,z) dx dy \right)^2} - 1, \qquad (11)$$

where $R$ is the radius of receiving aperture. We have taken $R = 5cm$ in our numerical study. Importantly, the aperture averaged scintillation is more stable than usually calculated on-axis SI our results, therefore, are more realistic. In figure (5), we have shown the aperture averaged SI for ring PVB. Figure 5(a) represents the aperture averaged SI in moderate turbulence for different topological charge $l$. It is noticed that with propagation the aperture averaged SI increases. On increasing the value of $l$ the aperture averaged SI of the beam becomes larger because larger $l$ makes outer wall of PVB thinner and this increases the aperture averaged SI. In strong turbulence, as depicted in figure 5(b), aperture averaged SI peaks at a particular distance of propagation and on further propagation it decreases and then saturates because of auto compensation effects [16].

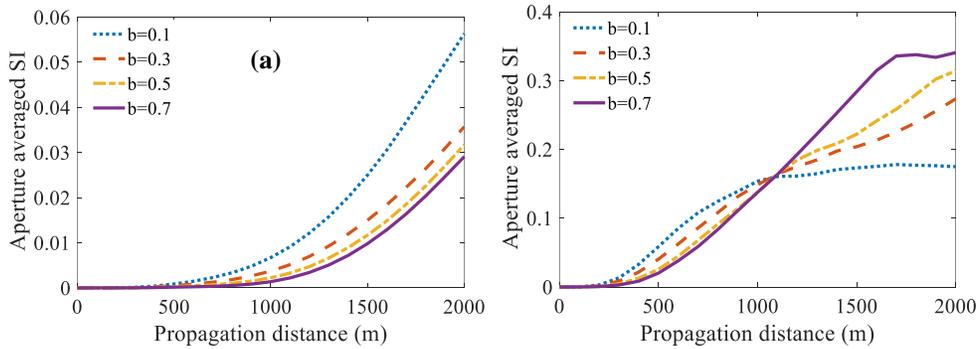

Fig. 6. Aperture averaged SI of ring PVB for $l = 1$ and different values of truncation factor: (a) $C_n^2 = 10^{-14} m^{-2/3}$ (b) $C_n^2 = 10^{-12} m^{-2/3}$.

In figure (6), we have studied the aperture averaged SI of ring PVB for different values of truncation factor. It is observed that the aperture averaged SI increases with propagation distance in moderate atmospheric turbulence. On increasing the value of truncation factor, aperture averaged SI reduces. In media with stronger turbulence (see figure 6 (b)), aperture averaged SI evolution becomes complex. For smaller value of the truncation factor (b=0.1) aperture averaged SI exhibits saturation effect. With increasing truncation factor the beam loses its self-healing capability and once the parameter is chosen to be very large the beam ceases to heal itself and therefore aperture averaged SI starts to increase with propagation.

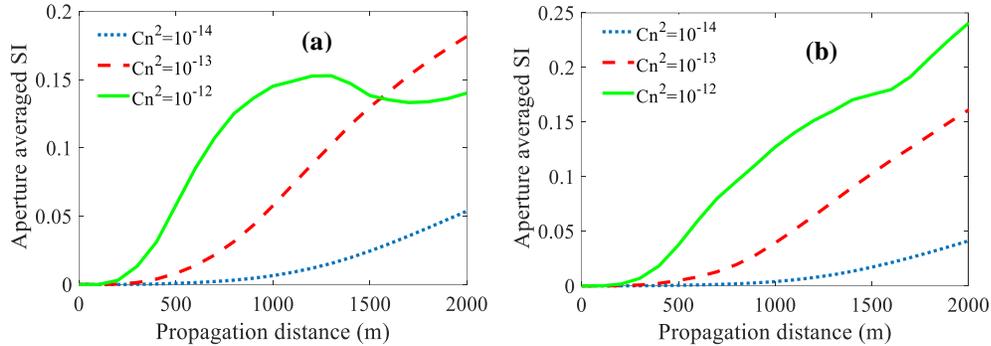

Fig. 7. Variation of aperture averaged SI with propagation for ring PVB beam with $l$ =1, (a) $b = 0.1$ and (b) $b = 0.3$ for different levels of turbulence.

In figure (7), we have varied the level of atmospheric turbulence and calculated the aperture averaged SI with propagation for $b = 0.1$ in figure 7(a) and $b = 0.3$ in figure 7(b). It is observed that in stronger turbulence, aperture averaged SI rises even at lower distances. Figure 7(b) depicts that stronger $b$ deteriorates aperture averaged SI.

### 4.3 Aperture averaged SI of spatially chirped ring PVB

In this section we have studied the effect of initial spatial chirp on aperture averaged SI of ring PVB whose mathematical expression is given by

$$E(r, \phi, 0) = Pe\left(\zeta_0, \frac{-r}{w_0}\right) exp\left[b\left(\frac{-r}{w_0}\right)\right] exp(il\phi) exp\left(ic\left(\frac{-r}{w_0}\right)^2\right), \tag{12}$$

where $c$ is the second order spatial chirp parameter and rest of the parameters are already defined in section (3).

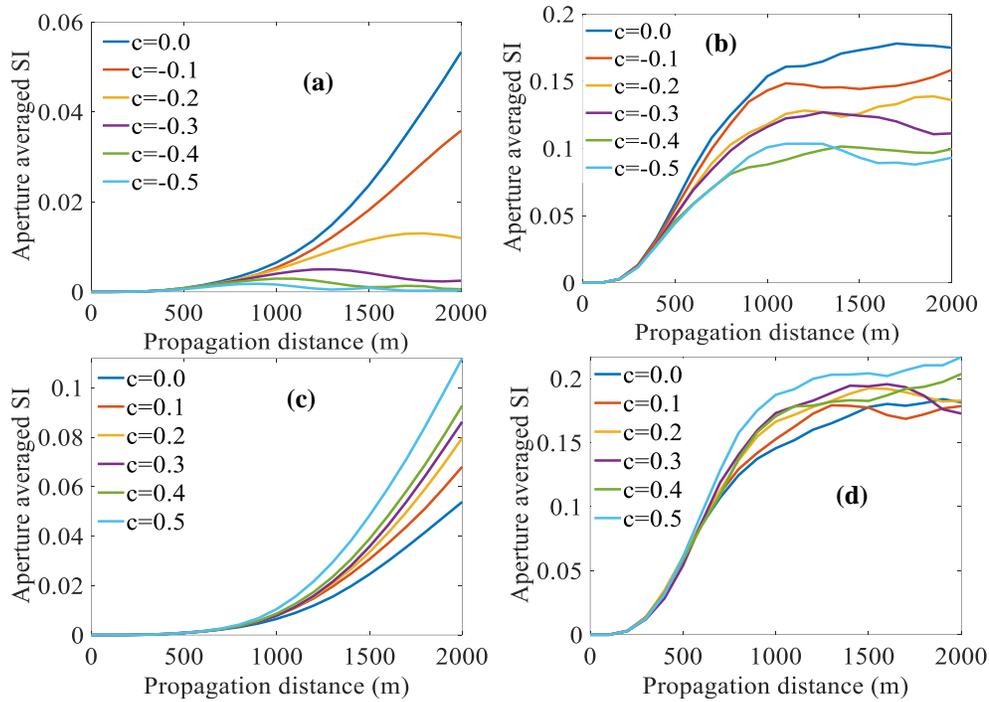

Fig. 8. Apertured average SI for chirped ring PVB with $l = 1$ and $b = 0.1$ for different values of chirp parameter. (a) & (c) $C_n^2 = 10^{-14} m^{-2/3}$ (b) & (d) $C_n^2 = 10^{-12} m^{-2/3}$.

In figure (8), we have calculated the aperture averaged SI for chirped ring PVB in moderate and strong turbulences. Spatial chirp embeds spherical phase in the input of ring PVB, which, depending upon the sign of the chirp, either makes the beam converging or diverging. In our case, introduction of the negative chirp at input enhances the convergence of the beam while positive chirp enhances the divergence of the beam. We observe that in moderate turbulence, the value of aperture averaged SI increases with propagation distance but as we decrease the value of chirp parameter (increase the value of the negative chirp) aperture averaged SI decreases. For very large negative chirp parameter, e.g., for $c = -0.4$ and $c = -0.5$ we see very small aperture averaged SI as depicted in figure 8(a). For beams with positive spatial chirp, aperture averaged SI increases with increase in chirp parameter as shown in figure (8c). In strong turbulence, (see figure 8(b)) aperture averaged SI with propagation first increases and then saturates. In this case also the larger value of the negative chirp parameter reduces the aperture averaged SI largely. Initial value of chirp affects the autofocusing properties of ring PVB. Figure 9 shows the evolution of ring PVB in free space with different initial chirp parameters. When there is no chirp (figure 9(b)), the beam focuses due to its auto focusing nature. With negative chirp (figure 9(a)), the beam converges quickly due to enhanced convergence. The second focus is mainly decided by second order chirp factor which makes the beam pass through a focusing lens [41]. Incorporation of positive chirp adds divergence to the beam and hence the beam quickly diverges as depicted in figure 9(c).

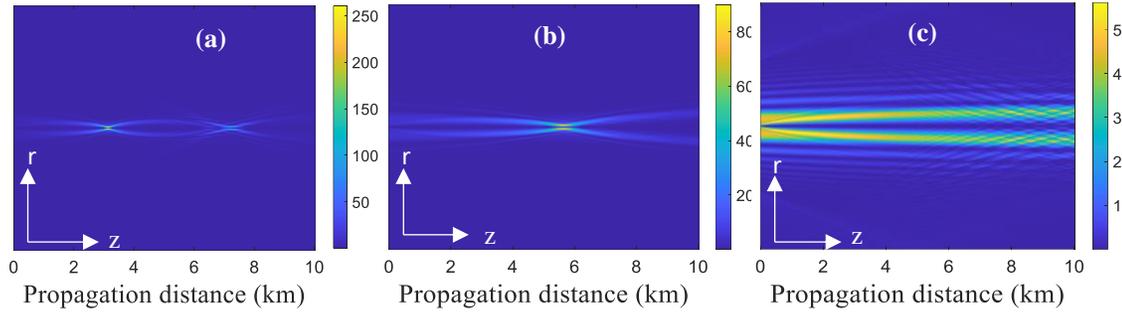

Fig. 9. Evolution of ring PVB with $l = 1$ and $b = 0.1$ in free space for different chirp parameter (a) $c = -0.2$ (b) $c = 0$ and (c) $c = 0.2$.

### 4.4 Scintillation of ring Airy vortex beam and ring PVB

In this section we have compared the aperture averaged SI of ring PVB and ring Airy vortex beam (AVB). We picked ring AVB for the comparison because till date, to the best of our knowledge, ring AVB has performed the best in atmospheric turbulence [14]. The mathematical expression of ring AVB reads as

$$E(r, \phi, 0) = Ai\left(\frac{R_0 - r}{w_0}\right) \exp\left(b\left(\frac{R_0 - r}{w_0}\right)\right) \exp(il\phi), \tag{13}$$

where Ai is airy function and $w_0$ is beam waist size $R_0$ is the radius of the primary ring and $b$ is the truncation factor. $R_0$ is taken to be zero in our simulations. It has been observed that in strong turbulence (see figure 10 (b)) ring PVB exhibits a better aperture averaged SI compared to ring AVB. Owing to multiple thinner rings PVB exhibits high SI at smaller distance as compared to that in ring AVB beam as shown in Figure 10(b). However, at larger distances, due to relatively stronger self-healing effect, ring PVB performs better and that leads to smaller aperture averaged SI in stronger turbulence. The better self-healing in the ring PVB owes its origin in its field structure. Figure 10(b) also depicts that the aperture averaged SIs for the two beams tend to saturate at larger distances. Although, in moderate turbulence (see fig. 10 (a)) ring AVB beam demonstrates better aperture averaged SI compared to ring PVB. Thus, based upon the above analysis, we can conclude that in stronger turbulence, in experimental scenarios, ring PVB may perform better as compared to ring AVB beam.

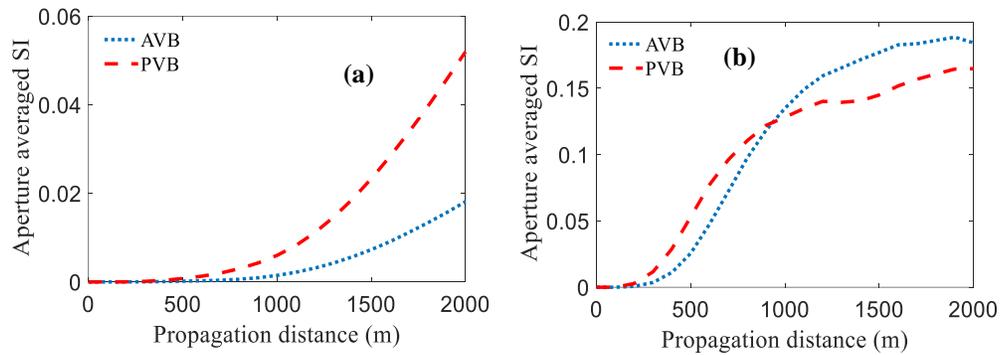

Fig. 10. Aperture averaged SI for $l = 1$ and $b = 0.1$ for ring AVB and ring PVB. (a) $C_n^2 = 10^{-14} \, m^{-2/3}$ (b) $C_n^2 = 10^{-12} m^{-2/3}$.

## 5. Beam wander of ring PVB and ring AVB

In this section we have computed beam wander for ring AVB and ring PVB and compared the two beams in both the moderate and strong turbulences based on their respective beam wanders. Random variation of beam centroid in turbulent atmosphere leads to the wandering of the beam beyond diffraction which is called beam wander. Scale size larger than beam diameter gives beam wander. The expression of beam wander that is valid in all turbulent conditions is given by [42, 43]

$$\langle r_c^2 \rangle = 4\pi k^2 W_{FS}^2 \int_0^L \int_0^\infty \kappa \phi_n(\kappa) exp(-\kappa^2 W_{LT}^2) \left\{1 - exp\left[\frac{-2L^2\kappa^2\left(1-\frac{z}{L}\right)^2}{k^2 W_{FS}^2}\right]\right\} d\kappa dz, \tag{14}$$

where $W_{LT}$ and $W_{FS}$ are long term spot sizes in turbulence and without turbulence respectively. $\phi_n(\kappa)$ is the PSD function. The expression of the $W_{LT}$ is the given by

$$W_{LT}(z) = \sqrt{2\frac{\int_{-\infty}^\infty \int_{-\infty}^\infty r^2 I(r,z)d^2r}{\int_{-\infty}^\infty \int_{-\infty}^\infty I(r,z)d^2r}}. \tag{15}$$

In our numerical simulation, we have calculated the long-term spot size $W_{LT}$ at different propagation distances with 500 independent realizations to get the averaged value of $I(r,z)$. While $W_{FS}$ is calculated at different propagation distances but with $C_n^2 = 0$. Substituting the values of $W_{LT}$ and $W_{FS}$ in equation (14) and considering modified Von Karman type PSD as mentioned in equation (4) and finally integrating equation (14) we get the beam wander of the optical beam. The dimensionless quantity $B_W = \frac{r_c^2}{W_{LT}^2}$ is more informative than the $r_c^2$ only. That's why we have calculated the $B_W$ for both ring PVB and ring AVB.

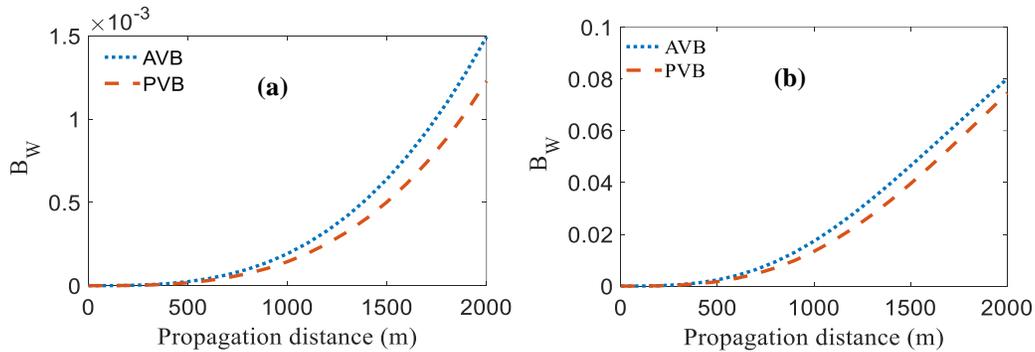

Fig. 11. Beam wander for $l = 1$ and $b = 0.1$ for ring AVB and ring PVB. (a) $C_n^2 = 10^{-14} m^{-2/3}$ (b) $C_n^2 = 10^{-12} m^{-2/3}$.

In figure (11), we see that with propagation distance beam wander increases and ring PVB performs somewhat better compared to ring AVB in both moderate and strong turbulences.

## 6. Conclusion

In summary, we have studied the propagation dynamics of ring PVB through atmospheric turbulence for different topological charges and truncation factors. We have studied the aperture averaged SI of ring PVB and found that on increasing the value of topological charge $l$, aperture averaged SI of ring PVB increases in moderate turbulence but shows saturation effect in strong atmospheric turbulence. Further, it has been observed that on increasing the value of truncation factor in moderate turbulence aperture averaged SI decreases while in strong turbulence the aperture averaged SI is found to increase with increasing value of the truncation factor. Moreover, ring PVB is found to outperform ring AVB in strong turbulence. Hence, our analysis showed that in real experiments ring PVB may perform better as compared to ring AVB. We have also studied spatially chirped ring PVB evolution dynamics in turbulent atmosphere. It is observed that spatial chirp strongly affects the aperture averaged SI. Our results show that a spatially chirped ring PVB with negative chirp parameter improves the aperture averaged SI. We have also computed beam wander of ring PVB and it has been found that ring PVB slightly outperforms ring AVB in both moderate and strong turbulences.

We believe that our study may find application in the FSO, remote sensing and space communication for better performance. In particular, the results of the work may improve the FSO and free space sensing.

## 7. Acknowledgement.

SS wishes to acknowledge the University Grant Commission (India) for the financial support. This work is supported by the research grant CRG/2022/007736 from SERB (India) and faculty initiation grant (IITR/SRIC/2386/FIG) from IIT Roorkee (India).

**Disclosure.** The authors declare no conflicts of interest.

**Data availability.** Data underlying the results presented in this paper are not publicly available at this time but may be obtained from the authors upon reasonable request.